\documentclass[conference]{IEEEtran}
\IEEEoverridecommandlockouts

\usepackage[subtle,tracking=normal]{savetrees} %

\usepackage{cite}
\usepackage{amsmath,amssymb,amsfonts}
\usepackage{algorithmic}
\usepackage{graphicx}
\usepackage{textcomp}
\usepackage[table,usenames,dvipsnames]{xcolor}
\definecolor{Gray}{gray}{0.9}
\usepackage{colortbl}
\usepackage{array}
\usepackage{booktabs}
\newcommand{\midsepremove}{\aboverulesep=0mm \belowrulesep=0mm}
\midsepremove
\newcommand{\midsepdefault}{\aboverulesep=0mm \belowrulesep=0mm}
\midsepdefault
\usepackage[hyphens]{url}
\usepackage{hyperref}

\usepackage[inline]{enumitem}
	\setlist[enumerate]{leftmargin=1em}%
	\setlist[itemize]{leftmargin=1em,label={$\bullet$}}%

\colorlet{myHighlight}{White!80!Green}
\usepackage{amsthm}
\usepackage{thmtools}

\usepackage{subcaption}

\newcommand{\subsub}[1]{\noindent\textbf{#1.} }

\usepackage{graphicx}
\usepackage{lipsum}
\usepackage{tcolorbox}
\usepackage[absolute,overlay]{textpos}

\begin{document}

\title{Exploring the Potential of Carbon-Aware Execution for Scientific Workflows  \\
}  
\author{
    \IEEEauthorblockN{
        Kathleen West\textsuperscript{a}, Fabian Lehmann\textsuperscript{b}, Vasilis Bountris\textsuperscript{b}, Ulf Leser\textsuperscript{b}, Yehia Elkhatib\textsuperscript{a}, Lauritz Thamsen\textsuperscript{a}
    }
    \IEEEauthorblockA{
        \textsuperscript{a}University of Glasgow, United Kingdom, \{firstname.lastname\}@glasgow.ac.uk \\
        \textsuperscript{b}Humboldt-Universität zu Berlin, Germany, \{firstname.lastname\}@informatik.hu-berlin.de
    }
}
\maketitle

\begin{textblock*}{\textwidth}(17.7mm,260mm)
    \begin{tcolorbox}[width=\textwidth, colback=gray!10, colframe=black, sharp corners, boxrule=0.5pt, boxsep=2pt]
        \centering \small \textbf{For the purpose of open access, we have applied a Creative Commons Attribution (CC BY) license to this version of our paper.}
    \end{tcolorbox}
\end{textblock*}

\begin{abstract}
Scientific workflows are widely used to automate scientific data analysis and often involve processing large quantities of data on compute clusters. As such, their execution tends to be long-running and resource intensive, leading to significant energy consumption and carbon emissions. 

Meanwhile, a wealth of carbon-aware computing methods have been proposed, yet little work has focused specifically on scientific workflows, even though they present a substantial opportunity for carbon-aware computing because they are inherently delay tolerant, efficiently interruptible, and highly scalable.

In this study, we demonstrate the potential for carbon-aware workflow execution. For this, we estimate the carbon footprint of two real-world Nextflow workflows executed on cluster infrastructure. We use a linear power model for energy consumption estimates and real-world average and marginal CI data for two regions.
We evaluate the impact of carbon-aware temporal shifting, pausing and resuming, and resource scaling. Our findings highlight significant potential for reducing emissions of workflows and workflow tasks.
\end{abstract} 


\section{Introduction}
Scientists across domains rely on increasingly large datasets and complex workflows to perform diverse tasks such as image processing, genome analysis, and material simulations. These workflows are pipelines composed of computational tasks. Systems like Nextflow~\cite{ditommasoNextflowEnablesReproducible2017a} allow for the execution and monitoring of scientific workflows on distributed cluster infrastructure. 

Scientific workflows often process vast quantities of data in parallel across numerous cluster nodes, and thus tend to be resource-intensive with runtimes spanning hours to weeks~\cite{diasDatacentricIterationDynamic2015}. This leads to significant energy consumption and carbon emissions.
For example, an Earth observation workflow~\cite{lehmann2021force} showed runtime variations (between 5 and 81 hours per execution) depending on available resources, highlighting the need to assess and optimize the execution of such workflows. 

Recent initiatives to enhance the sustainability of computing aim to align computational loads with the availability of low-carbon energy through carbon-aware computing~\cite{wiesnerLetWaitAwhile2021c, 10.1145/3626788, 9770383, 10.1145/3620678.3624644}. This alignment can be achieved by temporally shifting and scaling flexible compute workloads against energy signals like carbon intensity (CI), which is a measure of the emissions produced per kilowatt-hour ($kWh$) of electricity consumed.
There are two practically relevant CI signals: average and marginal. Average reflects the overall grid emissions, factoring in each energy source's relative share and emission rate. In contrast, marginal measures the emissions of the specific energy source meeting an additional load.
In many regions, both signals vary significantly due to intermittent renewable sources and demand fluctuations~\cite{wiesnerLetWaitAwhile2021c}. 
Temporal shifting involves scheduling applications to consume electricity when the CI is relatively low and to pause the workload otherwise~\cite{wiesnerLetWaitAwhile2021c, 9770383}. Resource scaling entails dynamically allocating resources to workloads based on the CI of electricity to make use of more resources when the CI is low, and to reduce demand when it is higher~\cite{10.1145/3626788}.  

While these methods demonstrate the potential of carbon-aware computing, no study to date has focused on carbon-aware scheduling and scaling of scientific workflows, despite them appearing particularly well-suited for carbon-aware execution owing to the following properties:  
\begin{itemize}
    \item \emph{Delay tolerance}: Many scientific workflows will not have strict deadlines (e.g. executing against a new dataset when it becomes available), so even if results will often be expected within certain time frames (e.g. a few days), there is flexibility for executions to be shifted.
    \item \emph{Interruptibility}: Workflows consist of tasks that typically exchange intermediate results between tasks using disks, allowing to pause execution temporarily and to execute subsequent tasks from persisted data when lower carbon energy becomes available again.
    \item \emph{Scalability}: Resource allocation can be adjusted so that tasks are executed on machines of varying scales and entire workflows are run on clusters of different sizes. 
    This enables the shaping of runtimes and resource usage against upcoming periods of low-carbon energy.
    \item \emph{Heterogeneity}: The tasks of a workflow can have varying resource demands, including possibly be CPU-intensive or I/O-intensive analysis steps, so especially energy-intensive tasks could utilize the lowest carbon energy available.
\end{itemize}
This paper explores the potential of carbon-aware execution for scientific workflows. 
Specifically, we evaluate the potential emission savings for two real-world Nextflow workflows.
We assess carbon-aware temporal shifting, with and without interruptions, and resource scaling at node and processor levels.
Moreover, we quantify emissions and possible savings by applying both average and marginal CI using commercial-grade data.
We explore potential emission savings by comparing the impact of adjusted executions, without considering CI forecasting errors or workflow performance estimation accuracy.
We share our simulation and results analysis code\footnote{\url{https://github.com/GlasgowC3lab/potential-carbon-aware-workflows}}. 

\section{Exploration of the Potential} 
We describe our experimental setup before exemplifying the potential reduction in emissions through applying carbon-aware computing methods.

\subsection{Experimental Setup}

\subsub{Scientific Workflows} 
We study two real-world workflows from Nextflow's community-curated nf-core library\footnote{\url{https://github.com/nf-core/}}~\cite{ewels2020nf} -- Chip-Seq and Rangeland. 
To minimize carbon emissions in our study, we rely on existing historical traces for entire workflow executions from~\cite{lehmannWOW2025}. 
Meanwhile, we executed the trimgalore\footnote{\url{https://github.com/FelixKrueger/TrimGalore}} workflow task on cluster and server resources. 

\subsub{Compute Resources}
The Chip-Seq workflow was executed on an eight-node cluster equipped with AMD EPYC 7282 processors with 128$GB$ RAM, while Rangeland was executed on an eight-node cluster equipped with Intel Xeon Silver 4314 processors with 256$GB$ RAM -- we executed workflow tasks on one of these nodes, alongside an edge server equipped with an Intel i7-10700T processor with 32$GB$ RAM. 

\subsub{Energy Consumption Estimation}
We estimate the carbon footprint using Ichnos~\cite{west2024ichnoscarbonfootprintestimator}, a tool we built for estimating the carbon footprint of Nextflow workflows from traces, which allows users to provide power models for the compute resources utilized. We used a linear power model to estimate energy consumption, translating this to carbon emissions with fine-grained CI data aligning with each workflow task's execution. 

\subsub{Carbon Intensity Data}
We performed all footprint estimations using average and marginal CI data sourced from the Electricity Maps Data Portal\footnote{\label{foot:ElectricityMaps}\url{https://www.electricitymaps.com/data-portal}} and WattTime\footnote{\label{foot:WattTime}\url{https://watttime.org/}}. 
We use both CI signals data at hourly intervals to make results directly comparable, except for experiments in Sections~\ref{subsec:compute-resource-experiment} and~\ref{subsec:freq-scaling}. 
This aligns with most carbon-aware computing research that uses the signals presented in Section~\ref{sec:RelatedWork}. While other metrics for renewable energy accounting exist (e.g. market-based measures such as RECs and PPAs), we did not focus on these as they not necessarily reflect low-carbon energy availability at a certain time and place.
While all scientific workflows and tasks were executed from 2023 to 2024, the latest average CI data available to us was from 2023. For workflows that occurred in 2024, we backdated the timestamps to 2023. 

\subsection{Baseline Carbon Footprint}
To establish baselines for subsequent experiments, we estimated the energy consumption of the selected workflows. With the Chip-Seq workflow consuming 35.16$kWh$, and the Rangeland workflow consuming 11.54$kWh$. We estimated the original carbon footprint for both workflows, in Table~\ref{table:entire-wf-shift}, using average and marginal CI data, based on the original execution times. 
We focus on operational carbon emissions and do not discuss embodied carbon emissions in this work.

As the compute cluster nodes were reserved solely for these workflows, we could also factor in the whole memory available on each node. We estimated the energy consumption for the `reserved memory' -- the full memory available on all utilized nodes over the workflow's execution -- finding that it accounted for 4--5\% of the overall estimated consumption (the sum of workflow and reserved memory consumption). Since workflows can also be executed on shared compute resources, we do not consider full node memory emissions in this way in our evaluation of carbon-aware shifting and scaling. 

\subsection{Entire Workflow Shifting}
\label{Subsec:DelayTolerance}
In our first experiment, the start time of each workflow's execution was adjusted by an hour, for every hour within a specified ``flexibility window'' to measure the potential reduction possible without adjusting the workflow's execution. 
We considered two windows: one of 2 days (24h before and after), and of 8 days (96h before and after). We considered start times before the original start time to ensure that low carbon windows were not missed, to fully illustrate the potential for carbon-aware shifting. 

The results are in Table \ref{table:entire-wf-shift}. 
We found that increasing the flexibility window could yield a greater reduction in the overall footprint for both workflows -- with the potential reduction largely dependant on when the original workflows were executed and the CI levels of the surrounding days. 
Our approach assumes an unrealistic knowledge of workflow runtimes and error-free CI forecasts. However, many schedulers are reliant on such signals~\cite{Topcuoglu_2002_HEFT,DURILLO2014221}, and tools are available to predict workflow task runtime and energy consumption with a low error~\cite{huangCloudProphetMachineLearningBased2024,BADER2024171}. Still, we emphasize that this study explores potential emissions reduction, but not how this potential can be achieved in practice.

\begin{table}[tb]
\caption{Reduction in carbon footprint shifting entire workflows. OrFP = Original Footprint}
\label{table:entire-wf-shift}
\centering
\rowcolors{3}{}{Gray}
\setlength{\tabcolsep}{4pt} %
\begin{tabular}{l|rrr|rrr}
 & \multicolumn{3}{c|}{Average CI} & \multicolumn{3}{c}{Marginal CI} \\
\toprule
 & \multicolumn{1}{c}{OrFP} & \multicolumn{2}{c|}{Reduction (\%)} & \multicolumn{1}{c}{OrFP} & \multicolumn{2}{c}{Reduction (\%)} \\
Workflow  & \multicolumn{1}{c}{ (gCO2e)} & in 48h & in 192h & \multicolumn{1}{c}{ (gCO2e)} & in 48h & in 192h \\
\midrule
Chip-Seq   & 19,386.45    & 49.16  & 63.98   & 27,615.39   & 5.94   & 66.21    \\
Rangeland & 3,926.96 & 32.60 & 48.99 & 10,075 & 5.89 & 7.67  \\
\bottomrule
\end{tabular}
\end{table}

\subsection{Interrupted Workflow Shifting} 
\label{Subsec:Interruptibility}
In our second experiment, we considered how workflows could be interrupted to exploit multiple shorter periods of low-carbon energy. While individual tasks cannot generally be paused and resumed, their start can be delayed without substantial overhead. As workflow systems like Nextflow use disk storage to exchange intermediate results, there will often be limited runtime overhead for reading the inputs of tasks from disks at a later point in time. 
The overhead of pausing and resuming entire workflow applications, hence, mainly stems from having to align task executions with multiple shorter periods of low-carbon energy availability, so that all tasks executed in a given time period finish fully within given periods.

For our experiment, we divided tasks from the original execution into hourly windows, to align them with multiple non-consecutive low-carbon CI windows. These windows contain two types of tasks:
\begin{enumerate*}[label=(\roman*)]
\item complete tasks that start and finish in the current hour;
\item partial tasks that start in the current hour but finish later. 
\end{enumerate*} 
All the partial tasks in a window that occurs before interruption will be executed in a later window. By interrupting workflow execution in this way, we add some overhead and estimate this by considering the longest partial task in a window (most delayed) -- as an upper bound of overhead. The overall overhead of interruptions is the sum of the overheads of individual windows for every interval where interruption occurred. 
We mapped the task execution windows to the lowest carbon intervals in a given flexibility window, in chronological order, to align with the workflow's original execution and data dependencies. 

\begin{figure}[h]
\centerline{\includegraphics[width=1\columnwidth,trim={0 0 0 0},clip]{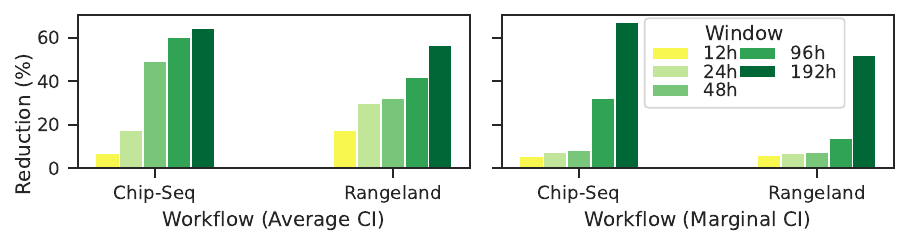}}
\caption{Reduction in carbon footprint of workflows using interrupted temporal shifting over 12--192 hour windows.}
\label{fig:ts-interrupt-reduction}
\end{figure} 

For Chip-Seq, a potential reduction of 49\% can be achieved with no overhead in the 48h window, and this improves to 64\% in the 192h window with a runtime overhead of $\approx$4\% -- using the average signal. 
For Rangeland, a potential reduction of 8\% can be achieved with a runtime overhead of $\approx$5\% in the 48h window, improving to a reduction of 52\% with a runtime overhead of $\approx$3\% in the 192h window -- using the marginal signal -- outperforming the entire workflow shifting. 
This exemplifies the potential for interrupted workflow shifting to offer greater reductions in carbon emissions over the same windows with a low overhead.
Furthermore, more sophisticated implementations of interrupted shifting, that divide workflows into shorter task execution windows, e.g. 15min instead of 60min, and more adaptive alignment of tasks, i.e. the most energy-intensive or longest-running tasks being aligned with windows of low CI electricity, could outperform entire workflow shifting even more significantly -- making better use of limited time before deadlines.

\subsection{Adjusting Compute Resources Used}  
\label{subsec:compute-resource-experiment}
\begin{figure}
\centering
\begin{subfigure}{0.45\columnwidth}
    \includegraphics[width=\textwidth]{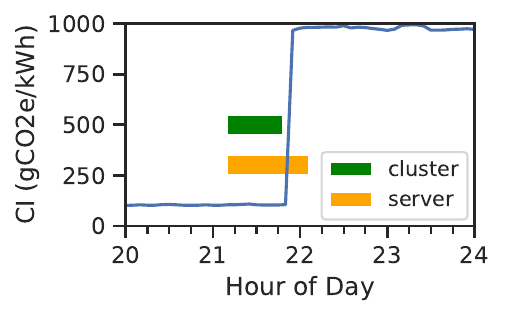}
    \caption{Resource assignment.}
    \label{fig:res-assignment}
\end{subfigure}
\hfill
\begin{subfigure}{0.45\columnwidth}
    \includegraphics[width=\textwidth]{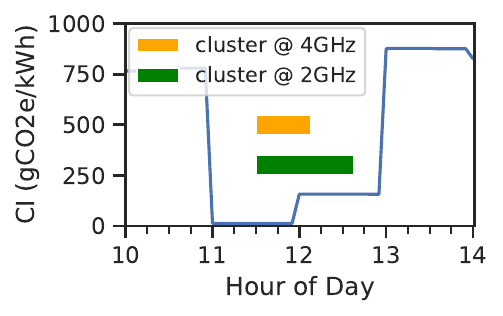}
    \caption{Frequency scaling.}
    \label{fig:freq-scaling}
\end{subfigure}
\caption{Impact of resource scaling on the alignment of trimgalore execution with marginal CI in California (a) and the Netherlands (b).}
\label{fig:figures}
\end{figure}

We explored the impact of choosing different nodes to execute the same individual workflow task, using 5-min marginal CI time-series data. 
We executed the trimgalore task on a cluster node, and an edge server, and recorded the runtime and energy consumption on the utilized resources.
In Fig.~\ref{fig:res-assignment}, we plot the marginal CI for the region of California on the 26th of November, 2023, showing selected executions of the trimgalore task, all starting at 21:10. At this time, the CI is below 200$gCO2e$, but it sharply increases around 21:50.
The task executed on the cluster ran for 33min, consuming 0.07$kWh$ and producing 37.9$gCO2e$. The task that ran on the edge server ran for 50min, consuming 0.02$kWh$ and producing 13.8$gCO2e$. 
We see that the smallest carbon footprint could be achieved by using a less powerful device, despite the cluster execution better fitting the low-carbon window. 

\subsection{Adjusting Processor Frequencies}  
\label{subsec:freq-scaling}
We also explored the impact of varying the processor frequency on the carbon emissions produced by running an individual task, using 5-min marginal CI time-series data. 
We executed the trimgalore task on the cluster node and varied the frequency, starting at 2.0GHz and increasing by 0.5GHz. We repeated this increase up until the frequency exceeded the device's maximum frequency. For all executions, we adjusted the start time to 11:30 on the 1st of July, 2023, using marginal CI data from the Netherlands, as shown in Fig.~\ref{fig:freq-scaling}, where we show selected executions of the task. We see that the CI rises twice, slightly at 12:00, then sharply at 13:00.

We could run the task on the cluster node running at 4.0$GHz$, which runs for 32min consuming 0.09$kWh$ to optimally fit the first half of the low-carbon window producing 4.0$gCO2e$. We could alternatively run the task on the same node at 2.0$GHz$, which runs for 62min consuming 0.10$kWh$ producing 9.5$gCO2e$. 

As shown, increasing the frequency of the node can reduce the time taken for the task to complete, and potentially the carbon emissions. So, we can `shape' a task by adjusting the processor's frequency, to fit tasks into low-carbon windows. 

\subsection{Adjusting Cluster Size for Entire Workflow Execution}
For this experiment, we used workflow traces for the execution of the Chip-Seq workflow at different scales on the compute cluster. In Table~\ref{table:cluster-scaling}, executing the workflow consumed approximately the same amount of energy at different scales, yet the makespan reduced as the number of nodes increased. We see that the execution on two nodes took 12 hours, while the execution on eight nodes took only 3 hours. These decreases in makespan lead to a reduction in the carbon footprint due to variations in the given CI data, especially for average CI. However, further reductions could be made by aligning shorter makespans optimally with low-carbon windows of electricity. 

\begin{table}[htb]
\caption{Impact of cluster sizes on the carbon footprint of Chip-Seq executions}
\label{table:cluster-scaling}
\centering
\rowcolors{3}{}{Gray}
\begin{tabular}{lrrrr}
\toprule
\# & makespan & energy & Avg. emissions & Marg. emissions \\
nodes & (h)  & (kWh) & (gCO2e) & (gCO2e) \\
\midrule
2     & 11.84 & 34.13 & 11,862.16 & 25,387.30 \\
4     & 5.97 & 34.39 & 9,616.61 & 25,968.88 \\
8     & 3.13 & 34.17 & 6,817.35 & 23,639.88 \\
\bottomrule
\end{tabular}
\end{table}

\section{Related Work} 
\label{sec:RelatedWork}

\noindent\textbf{Temporal Shifting}. 
Various works propose to schedule or interrupt flexible workloads to only consume electricity when CI is low, thereby reducing carbon emissions~\cite{wiesnerLetWaitAwhile2021c, 9770383}. Other studies aim to only leverage excess renewable energy, similarly delaying or scheduling applications to align with this energy and spare compute capacity~\cite{10.1007/978-3-031-12597-3_14, 10.1145/3632775.3639589}. 
None of these works specifically optimizes workflow execution.

\noindent\textbf{Resource Scaling}.
Carbon-aware resource scaling dynamically allocates more resources when CI is low and reduces demand when it is higher -- either performing horizontal scaling based on one-time offline profiling~\cite{10.1145/3626788}, or applying vertical scaling to limit the carbon emissions rate of containerized applications~\cite{10.1145/3620678.3624644}, which is not directly applicable to applications composed out of connected tasks like scientific workflows. 

\noindent\textbf{Other Carbon-Aware Computing Techniques}.
Other work has yielded carbon-aware computing techniques and tools to prioritize sustainability for data centres or to enable applications to control how they use renewable energy~\cite{10.1145/3575693.3575754, chien2023reducing}. 
Previous work considering specifically workflows has not gone beyond location-based load shifting~\cite{9298863}, or example use of carbon-aware time shifting~\cite{BADER2024171}. In addition, none of these techniques specifically exploits the characteristics of workflows for carbon-aware execution. 

\section{Conclusion} 
In this paper, we highlighted the potential of carbon-aware execution for scientific workflows. To begin, we estimated the operational carbon footprint from running two real-world workflows; they produced 3.9--19.4$kg$ of carbon emissions. 
Using these estimates as baselines, we first assessed carbon-aware time shifting, finding that this could reduce the footprint of Chip-Seq by up to 64\% using average CI and 66\% using marginal CI data. We also explored resource scaling as a means to either align the execution of individual workflow tasks to upcoming low-carbon energy availability windows or to assign more compute nodes to reduce the runtime of entire workflows -- potentially reducing the footprint of Chip-Seq by 42.5\% using average CI data.
Although the studied workflows have smaller footprints than those running for thousands of core hours, their impact is still significant.
Moreover, we see no reason why the potential identified should not be extrapolated to larger workflows, given also larger infrastructures and longer flexibility windows for executing workflows, since the workflow properties we highlight -- delay tolerance, interruptibility, scalability, and heterogeneity -- should be widely applicable.

\section*{Acknowledgments}

This work was supported by the Engineering and Physical Sciences Research Council under grant number UKRI154. 
We also gratefully acknowledge the sources of electricity grid data: NESO Open Data and Electricity Maps historical data for average carbon intensity, as well as the marginal operating emission rates calculated by WattTime.

\bibliography{refs}

\end{document}